\documentstyle[procl, twoside]{article}
\input{psfig.sty}

\def\Journal#1#2#3#4{(#1) {#2} {\bf #3}, #4} 
 
\def\AaA{\em Astron.\ \& Astrophys.} 
\def\AaAS{\em Astron.\ \& Astrophys.\ Suppl.\ Ser.}
\def\AJ{\em Astron.~J.} 
\def\ApJ{\em Astrophys.~J.}

\def\Nat{\em Nature\/}

\begin{document}

\markboth{PdBI observations of GMCs in M\,31}{N. Neininger}
\thispagestyle{plain}
\title{Molecular cloud complexes in detail: Interferometric
observations of GMCs in M\,31} 

\author{N. Neininger \\
(in collaboration with M. Gu\'elin, R. Lucas and S. Muller, IRAM
Grenoble)
}

\address{Radioastronomisches Institut der Universit\"at Bonn,\\
Auf dem H\"ugel 71, 53121 Bonn, Germany}

\maketitle

\abstract{
We have accumulated $^{12}$CO(1-0) and (2-1) data of several GMCs in
M\,31 using the Plateau de Bure Interferometer. The sample covers a
range of 5 to 18 kpc galactocentric distance and various physical
conditions.  The spatial resolution attains values down to 3\,pc. All
GMCs investigated have been resolved into several components, also
seemingly quiescent clouds and not only the cases where already the
survey positions show obvious multiple-peaked spectra. Such
kinematically disjunct emission is however spatially coincident in the
small volume of the interferometer beam in M\,31, which strongly
favours local effects as the cause of the velocity splitting. This is
well consistent with the observed absence of strong streaming motions
in the survey data. The nature of the sometimes strong separation in
velocity space needs further investigation. The complete data set
(i.e.\ including the large-scale survey) yields a uniquely complete
view of the molecular gas which allows to investigate the conditions
for star formation in detail and helps to establish guide lines for the
derivation of the properties of molecular gas.  }

\section{The Framework}
We have started a complete high-resolution survey of the $^{12}$CO in
M\,31 in 1995 in order to investigate the large- and small-scale
properties of the molecular gas with the highest possible angular
resolution and sensitivity (see the contribution of M. Gu\'elin, this
volume). One aim of this survey is to pinpoint the giant molecular
cloud complexes (GMCs) in this close-by spiral galaxy for follow-up
observations. Such a precise large-scale map is mandatory as a basis
for further investigation because the arm-interarm contrast of the
molecular gas has shown to be very high with a rather low general
``filling factor'' of the projected disk. The information from the
survey now allows e.g.\ to select pointings for weaker tracers or to
choose regions that merit detailed investigations with high angular
resolution with the Plateau de Bure interferometer (PdBI).

M\,31 is a quiescent galaxy with a low star forming rate, showing
well-defined spiral arm segments; however, they resisted up to now
against being put into a single global pattern. The high inclination
makes a kinematical analysis on the basis of the H{\sc i} data
difficult (see e.g.\ Braun 1991), but rather large streaming motions
had been derived in the north-eastern part from CO observations (Ryden
\& Stark 1986). The mapped areas of all such earlier work were however
generally by far too small to derive general properties of the
molecular gas. Our survey data now clearly show that the magnitude of
non-circular motions is only of the order of 10\,km\,s$^{-1}$ which is
the typical line width of the observed molecular gas. The total range
we observed is from about 4\,km\,s$^{-1}$ to something like
15\,km\,s$^{-1}$ for individual spectra. However, we found {\em strong
small-scale} disturbances at many places.

Typical examples are double- or multiple-component spectra, broad
lines and short-range spatial variations (cf.\ Fig.~1). Unlike similar
observations in the Milky Way, the location of the molecular clouds
with respect to other constituents of the ISM can be determined to a
much higher precision -- in particular, the distance ambiguities in
the interpretation are virtually absent in the M\,31 CO data.
Therefrom we are led to conclude that we are not observing chance
line-of-sight coincidences of spatially separated clouds, but true
local effects. The sizes of such regions are of the order of a few
primary beams of the PdBI, hence ideally suited for a detailed
observation.

\section{The sample}
During the observations for the survey a large number of interesting
regions showed up throughout the disk of M\,31. For more detailed
investigations some of them were subsequently covered at a denser
sampling in both, the $^{12}$CO(1-0) and (2-1) transitions, with
typical map sizes of a few ten arcminutes squared. For the PdBI
observations, we choose GMCs or GMC groups that can be covered with a
not too large mosaic while representing a broad spectrum of the cloud
complexes found. The innermost cloud in the sample is located at a
radius of $\sim5$\,kpc, the outermost group at $\sim18$\,kpc -- this
spans the whole range over which we detected CO emission in the
survey. In addition, we choose other GMCs with various morphologies in
different environments.

Up to now, our sample consists of six regions covered with 2- to
9-field mosaics -- mainly in the south-western part of the galaxy, but
also covering the prominent association of molecular gas on the
northern major axis at a distance of 5\,kpc from the centre.

\section{A remarkable pair}
Among the first objects studied in this way were two neighbouring
complexes in the main ``ring'' of emission, separated by less than a
kpc (at the location of the dust clouds D47 and D84; see Hodge
1981). One of them shows a relatively strong ($\sim 1\,$K, $\Delta
v\sim22\,$km\,s$^{-1}$), single peak in the survey (D84), the other
two narrow peaks separated by $\sim 20\,$km\,s$^{-1}$ (D47) -- see
Fig.~1. They have not only the same distance to the centre, they lie
even within the same continuous ridge of emission. Nevertheless, there
must be a substantial difference in the local conditions. The H{\sc i}
map (Brinks \& Shane 1984) shows many places with such multiple
spectra, but in view of the thickness of the atomic gas disk, the
inclination of M\,31 and possible distortions (Braun 1991) they can be
explained as multiple arms in the same line of sight (Berkhuijsen et
al.\ 1993).

\begin{figure}
\vbox{
\psfig{file=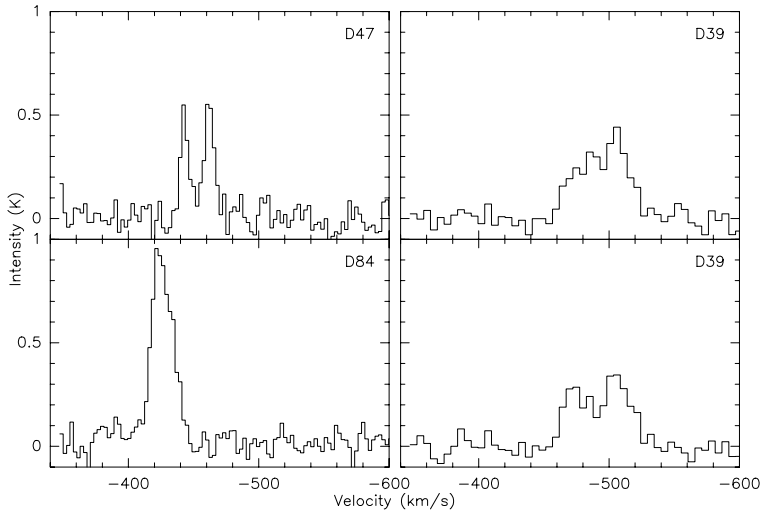,width=10cm,angle=270}\vspace{-50mm}}
\hfill\parbox[b]{5cm}{
\caption
{Four example spectra from the SW part of our survey; all boxes have
the same dimensions.  The dust clouds D47 and D84 are located about
13\,kpc west from the centre in the same continuous ridge of molecular
emission, separated by less than 1\,kpc. The two spectra of D39
(located in the south-west, close to the major axis) are separated by
only $10.6''$, which is less than half a beam; note their large width
and the kinematical differences already on this small scale. Other
GMCs on the major axis do not show such variations.\hspace{\fill} } 
\label{ho-figspek}
} 
\end{figure}

This explanation is not applicable here, however: the disk of
molecular gas is much thinner than the H{\sc i} disk and the observed
large-scale structure clearly excludes the existence of multiple arms
within one line of sight. The same holds for the spectra in the D39
region  (which are also multiple-peaked or unusually broad) since it
is located close to the major axis. Fortunately, a wealth of
complementary data exists for M\,31 which allows to look for possible
influences. In particular, we produced an overlay with the H$\alpha$
map from Devereux et al.\ (1994) and compared the positions with other
signs of activity such as stellar associations.

\begin{figure}
\vbox{
\psfig{file=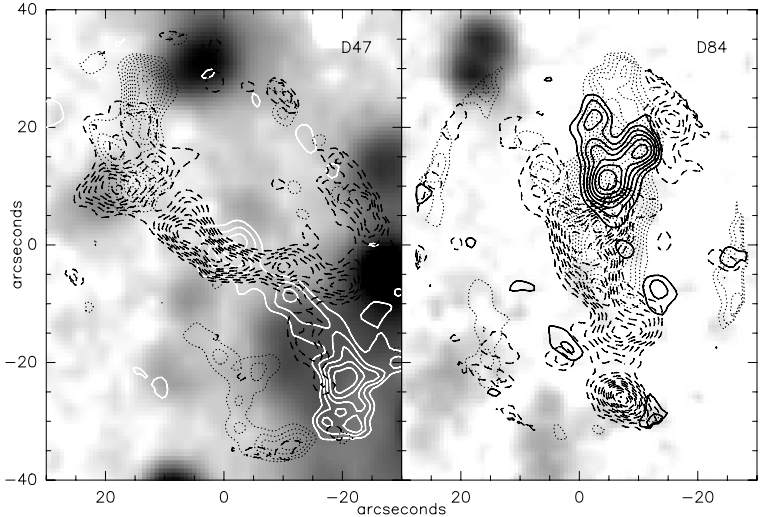,width=10cm,angle=270}\vspace{-70mm}}
\hfill\parbox[b]{5cm}{
\caption
{The CO emission of D47 and D84 compared with the H$\alpha$ emission
(darker tones mark higher intensities; from
Devereux et al.\ 1994). Three representative channels each are shown
as contours (solid: low, dotted: intermediate and dashed: high
velocity) as in Fig.\,2 of Neininger (1999). The primary-beam
correction of the mosaic enhances the noise along
its border.  The molecular gas of D47 is concentrated in
disjunct filaments, whereas D84 consists of kinematically coherent
clumps. Obviously, D84 is located in a quiescent environment, whereas
the GMC in D47 lies on the border of a strong H{\sc ii} region. This
bubble (\#281 in Pellet et al.\ 1978) is one of the
brightest H{\sc ii} regions in M\,31. \hspace{\fill} }
\label{ho-figcoha}
} 
\end{figure}

For these three cases we can deduce an easy explanation: the presence
of multiple-peaked spectra corresponds to enhanced star formation
activity in the vicinity of the GMC (cf.\ Fig.\,2). The region of D39
is known to host several OB associations and is adjacent to one of the
biggest H{\sc i} holes of M\,31 (Brinks \& Bajaja 1986, \#8). Close to
D47 we find the bubble-shaped H{\sc ii} region 281 (Pellet et al.\
1978) which is one of the ten brightest in M\,31 and {\em the}
brightest of this morphology. The CO ridge tends to avoid the bubble,
but the filaments visible in Fig.\,\ref{ho-figcoha} are located right
at the border of the ionized gas. D84, on the other hand, is rather
far from any sign of activity -- neither strong H$\alpha$ nor large
H{\sc i} bubbles are found here. This leads to the suggestion that
such local energy sources are responsible for line splittings or
multiple components in the CO emission. The nature of the interaction
needs however further investigation, as well as the r\^ole of the
individual constituents (stars, ionized, neutral and molecular gas).

\section{GMCs in detail}

To further investigate the structure of the GMCs, we obtained data of
an inner cloud complex (in the D153 dust cloud, galactocentric radius
$\sim5$\,kpc) at the highest possible angular resolution. We used the
``A''-configuration of the PdBI which allows sub-arcsecond resolution
in the CO(2-1) transition. At the distance of M\,31, this corresponds
to about 4\,pc in linear scale.

\begin{figure}
\vbox{
\psfig{file=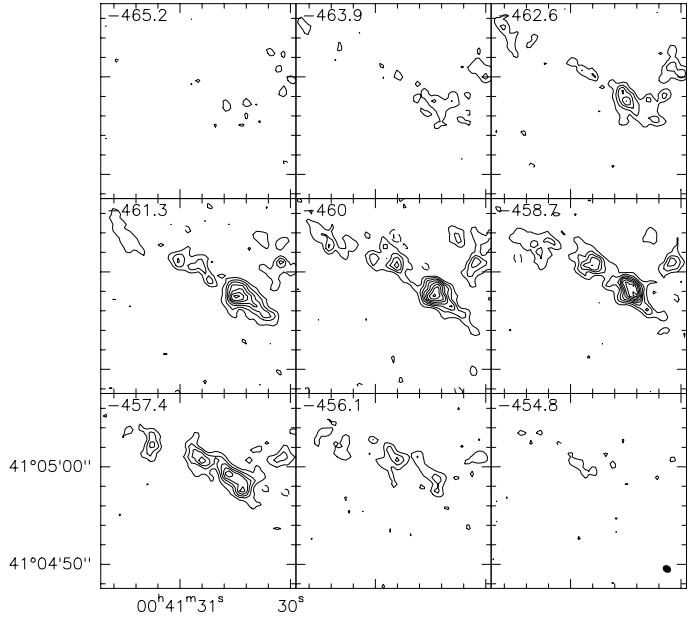,width=9cm,angle=270}\vspace{-68mm}}
\hfill\parbox[b]{5cm}{
\caption
{Channel maps of the $^{12}$CO(2-1) emission of a GMC at 5\,kpc from
the centre (dust cloud D153). Contours are spaced by 50\,mJy/beam
(FWHM $0.9''\times0.74'' \simeq 3.95\,{\rm pc}\times3.25\,{\rm pc}$,
indicated in the lower right corner). The velocity channel of each
field is marked in the upper left corner. J2000.0 coordinates are
given with the lower left field.
\hspace{1em} Although the structure is quite coherent in velocity
space, it has significant spatial substructure. At least four
individual clumps can be identified easily; no obvious kinematical or
spatial structure is indicated in the corresponding survey data at
$23''$ resolution. Even for this quiescent cloud the virial masses
determined from the two data sets differ significantly. \hspace{\fill}
} \label{ho-fig21in}
}
\end{figure}

Although the survey data show an almost structure-less blob at this
position, it is clear from Fig.\,\ref{ho-fig21in} that there is
significant substructure in this complex. What looks like a
well-defined entity in the survey at $23''$ resolution ($\sim 90$\,pc)
is resolved into at least four individual clumps at the higher angular
resolution.  In this case, the basic elements seem to be clumps
(similar to D84) rather than filaments as seen in D47. It has to be
checked, however, whether this is a true distinction depending on the
activity of the particular region or an effect of the lower resolution
of the D47 and D84 data. 

A comparison between the data at different resolutions allows us to
check mass determinations on the basis of the virial theorem. Some
calculations on the basis of the low-resolution data turned out to be
wrong by more than a factor of ten. In the D39 region, this virial
mass is up to a factor of 100 higher than the mass derived from
high-resolution data or the optically thin mm dust emission. Hence, a
systematic study of the GMC properties at various resolutions and
comparing different approaches is essential to obtain better
guidelines for the derivation of such fundamental values.

\section{(Preliminary) conclusions}

A statistical analysis of the GMCs in the survey as well as a more
detailed investigation of the properties of the clumps in the PdBI
data is under way. Already the few rather compact cloud complexes
investigated thus far hint however at important implications for the
properties of molecular gas agglomerations and their analysis: \\[1ex]
$\ast$ CO spectra in M\,31 showing multiple lines are usually confined
to small regions. \\
$\ast$ The thin molecular disk and the high spatial resolution
indicate a true local phenomenon. \\ 
$\ast$ Many of these spectra are found near tracers of locally 
enhanced star formation. \\
$\ast$ The internal structure of such GMCs consists of well separated
filaments or clumps. \\ 
$\ast$ The velocity separation of the components is significantly
greater than the streaming motions. \\ 
$\ast$ Even seemingly quiescent clouds may consist of individual
clumps which may afflict the derivation \hspace*{1em}of the parameters
of the cloud, such as mass or temperature. \\[1ex]
 This is the first time that such a wide range of
scales of the molecular gas emission can be investigated and searched
for clues about its state, evolution into stars and mutual
interactions with the other constituents of the ISM.

\section*{Acknowledgements} 
 
It is a pleasure to thank the PdBI staff and astronomers for the
excellently performed observations. However, the tragical accidents
stopped the completion of the high-resolution data set (shown in
Fig.\,\ref{ho-fig21in}) among many other projects. This work is
therefore dedicated to the memory of the victims.

A lot of work for the survey was done by Ph.\ Hoernes and in
particular by Ch.\ Nieten, who set up (and efficiently uses) the
present data reduction and analysis routines capable to handle the
huge amounts of information fast enough. 
 
\section*{References}\noindent
 
\references 

Berkhuijsen E. M., Bajaja E., Beck, R,: \Journal{1993}{\AaA}{279}{359}

Braun R.: \Journal{1991}{\AaA}{372}{54}

Brinks E., Bajaja E.: \Journal{1986}{\AaAS}{169}{14} 

Brinks E., Shane W.W.: \Journal{1984}{\AaAS}{55}{179} 

Devereux N. A., Price R., Wells L. A., Duric N.:
\Journal{1994}{\AJ}{108}{1667}

Hodge P. W.: (1981), {\it Atlas of the Andromeda Galaxy}, University of
Washington Press, Seattle and London

Neininger N.: (1999), in V. Ossenkopf et al.\ (eds.): {\it The Physics and
Chemistry of the Interstellar Medium} 
GCA-Verlag, Herdecke,  p.\ 42

Neininger N., Gu\'elin M., Ungerechts H., Lucas R., Wielebinski, R.:
\Journal{1998}{\Nat}{395}{871} 

Pellet A., Astier N., Viale A., et al.:
\Journal{1978}{\AaAS}{31}{439} 

Ryden B. S., Stark A.A.: \Journal{1986}{\ApJ}{305}{823}
\end{document}